\documentclass{aa}
\usepackage{graphicx,natbib}
\def\bp{$\beta$\,Picto\-ris}
\def\hra{HR\,4796\,A}
\def\hrb{HR\,4796\,B}
\def\hrd{HR\,4796\,D}
\def\cs{\mbox{circumstellar}}

\def\h{\hfill\break}
\newcommand{\ma}[1]{\mathrm{#1}}
\newcommand{\uma}[1]{^{\mathrm{#1}}}
\newcommand{\dma}[1]{_{\mathrm{#1}}}
\begin{document}
\thesaurus{08 (19.42.1 \hra\ ; 19.25.1; 04.01.1; 05.01.1; 19.94.1)}
           %
           %
\authorrunning{J.C. Augereau et al.}
\titlerunning{On the \hra\ \cs\ disk}
\title{On the \hra\ \cs\ disk \thanks{Based on observations collected at the
European Southern Observatory, La Silla, Chile}}
\author{J.C. Augereau\inst{1} \and A.M. Lagrange\inst{1} \and
D. Mouillet\inst{1} \and J.C.B. Papaloizou\inst{2} \and P.A. Grorod\inst{1}}
\offprints{J.C. Augereau}
\mail{augereau@obs.ujf-grenoble.fr}
\institute{
Laboratoire d'Astrophysique de l'Observatoire de Grenoble, Universit\'e J.
Fourier / CNRS, B.P. 53, F-38041 Grenoble Cedex 9, France
\and Astronomy Unit, School of Mathematical Sciences, Queen Mary \& Westfield
College, Mile End Road, London E1 4NS, UK}
\date{Received {...}; accepted {...}}
\maketitle
\begin{abstract}
We investigate in details the properties of the disk surrounding the 8\,Myr
old star \hra, one of the few stars bringing precious clues to better
understand the scenario which lead to planetary system formation.
We propose a model able to reproduce all the available
observations~: the full spectral energy distribution from the
mid-infrared to the millimeter wavelengths, resolved scattered light
and thermal emission observations.
We show that the \cs\ matter splits into two dust components~: a cold annulus,
peaked at 70\,AU from the star, made of ISM-like grains (amorphous composition,
po\-ro\-sity $\sim$ 0.6) larger than 10\,$\mu$m and a population of hot dust 
close to the star (at about 9\,AU) made of comet-like grains (crystalline
composition, porosity $\sim$ 0.97). Both dust populations are highly
collisional and the grain size distribution in the cold annulus is found to be
cut-off by radiation pressure. At 70\,AU, bodies as large as a few meters are
required to fit the data leading to a minimum disk mass of a few Earth masses
and to a gas to dust ratio less than $1$.
We discuss aftewards some implications on the disk structure and
effects of larger bodies.
\end{abstract}
%
%
%
%
%
%
\section{Introduction}
\hra\ is one of the A type stars showing a high infrared excess
($L\dma{disk}$ / $L_{*} = 5.10^{-3}$) due to the thermal emission from \cs\
dust. The excess is twice that of \bp, the best
studied case among Vega-like stars so far. The star age, 8$\pm$2\,Myr
\citep{stau95}, is about 5 to 15 times less than \bp\ implying that \hra\ is
tracing an evolutionary phasis prior to that of \bp.

\citet{koe98} and \citet{jaya98} resolved first the \cs\ dust at
thermal infrared wavelengths ($\lambda=10 - 20\,\mu$m) recently followed by
\citet{sch99} in scattered light ($\lambda=1.1$ and $1.6\,\mu$m).
Interestingly, the disk lies nearly in the direction to \hrb, a physically
bound companion of \hra\ \citep{jura93},
located 7.7" away, \textit{i.e.} 515\,AU in  projected distance according to
the \textit{Hipparcos} star distance $d_{*}=67.1_{-3.4}^{+3.5}$\,pc.

According to the resolved images, the disk is extending only a few tens of AU
outside the peak of the dust distribution assessed to about 70\,AU
\citep{sch99}.
Outer truncation of circumstellar disks within a binary system
has been predicted at typical distances of 1/3 -- 1/2 of 
the binary separation \citep{pap77,arty94}. However, the impact of the
companion \hrb\ on the disk extension is still unclear.

In the inner part of the disk, \citet{jura95} first suggested a depletion of
dust close to the star so as to reproduce the 110\,K color temperature deduced
from IRAS data.
The comparison between resolved mid-IR images and first order modeling
(see below) confirmed the necessity of an inner hole
in the disk at 55$\pm$15\,AU from the star for \citet{koe98} and 60$\pm$20\,AU
for \citet{jaya98}.

According to \cite{koe98}, a second population of hotter grains may however
lie closer to the star (inside the inner hole). Located at distances
similar to those of the zodiacal dust in our Solar System, this dust population
would be responsible for both the 12.5\,$\mu$m detected excess and for a
faint emission at 20.8\,$\mu$m (less than 10\% of the total flux at the
wavelenght) in excess of the resolved disk and centered on the star.

The grains properties are so far poorly constrained.
According to \citet{jura95}, grains smaller than about 3\,$\mu$m are blown
outward by radiation pressure. 
Considering the Poynting-Robertson effect, they found that the grains are
probably larger than 40\,$\mu$m under the assumption of a 40\,AU inner hole.
\citet{koe98} used thermal absorption/emission laws for the
grains following the model proposed by \citet{bac92} for \bp~: $Q\dma{abs}$ is
constant for $\lambda<\lambda_0$ and proportionnal to $\left(\lambda /
\lambda_0\right) ^{-1}$ otherwise, the parameter $\lambda_0$ is expected
to be related to an effectif grain size $a_0$ in the disk.
Although \citet{koe98} constrain $\lambda_0$ to 25$\pm$15\,$\mu$m,
they do not specify the ratio $\frac{\lambda_0}{a_0}$ in the case of
\hra. This ratio could vary by a factor 40 between the lower and the higher
value \citep{bac92}.

Before beeing imaged in the mid-IR spectral range, no optical or near-IR
observations succesfully resolved the disk around \hra, if we except for a
slight asymmetry in the coronographic adaptive optics system images
performed with the ESO adaptive optics system in K' band \citet{mou97}.
Because of the low signal to noise ratio in this data, these results were not
presented as a positive detection. Knowing the position angle of the
disk, we reduced again these observations and we now marginally detect the
disk. After a brief summary of thermal available data, we present in section
\ref{secdata} the newly reduced scattered light images and compare then with
more recent and better signal to noise ratio scattered ligth images
\citep{sch99}.

As data become actually more numerous, we wish to better constrain
the grain properties in the \hra\ disk.
Sophisticated grains models developed by \citet{gre72}
succeeded both in reproducing interstellar observations
(extinction, polarization \citep{li97}) and in fitting the shape of the SED
of the disk surounding \bp\ \citep{li98,pan97}.
In addition, \citet{li98} proposed a link between ISM particles and \cs\
evolved environments assuming that dust grains in the \bp\ disk are
of cometary origin (see also \citet{bac93,lec96}) and could be fluffy
aggregates of primitive interstellar dust.

In this paper, we adopt a similar approach for the grains in the \hra\ disk.
We describe section \ref{secmodel} the disk model assumptions and try to fit
(sections \ref{single} and \ref{double}) the Spectral Energy Distribution
(hereafter SED) to derive some physical and chemical grain properties such as
typical size, porosity, presence of ice and finally to estimate if the grains
are more similar to interstellar dust grains or to comet-like grains.

In addition, we test whether the presence of a second population responsible
for the 10\,$\mu$m excesses is necessary or not.
Given the constraints on the grain distribution derived from the SED fitting,
we try to reproduce the thermal and scattered light resolved images (section
\ref{images}). We finally discuss section \ref{discuss} the implications on
the disk dynamics.
%
%
%
%
%
%
\section{Available data}
\label{secdata}
\subsection{Spectral Energy Distribution}
The available photometric data for the \hra\ disk are summarized in Table
\ref{data}. They include color-corrected IRAS measurements, IRTF bolometer
measurements \citep{faj98}, Keck/Mirlin \citep{koe98} and Cerro Tololo/ OSCIR
\citep{jaya98} images and the IRTF bolometer datum obtained by \citet{jura93}
at 20\,$\mu$m. They also include the available submillimeter data~:
JCMT upper limit at 800\,$\mu$m \citep{jura95} and recent SCUBA
measurements at 450\,$\mu$m and 850\,$\mu$m \citep{grea99}.
\begin{table}[!h] 
\begin{center}
\begin{tabular}[h]{rccc}
\hline \hline
$\lambda$\,\,\,\, & Excess & Uncertainty &  \\
$[\mu$m$]$ & $\Phi\dma{obs}$ $[$Jy$]$ & $\Delta\Phi\dma{obs}$ $[$Jy$]$ & Reference \\
\hline \hline
4.80 & no excess &  & Fajardo-A. et al. (1998) \\
7.80 & 0.067 & 0.037 & Fajardo-A. et al. (1998) \\
9.80 & no excess &   & Fajardo-A. et al. (1998) \\
10.1 & 0.087 & 0.026 & Fajardo-A. et al. (1998) \\
10.3 & 0.057 & 0.024 & Fajardo-A. et al. (1998) \\
$^*$11.6 & 0.086 & 0.070 & Fajardo-A. et al. (1998) \\
$^*$12.0 & 0.173 & 0.028 & IRAS \\
$^*$12.5 & 0.101 & 0.018 & \citet{koe98} \\
$^*$12.5 & 0.133 & 0.027 & Fajardo-A. et al. (1998) \\
$^{*}$18.2 & 1.100 & 0.150 & \citet{jaya98} \\
20.0 & 1.860 &  unknown  & \citet{jura93} \\
$^{*}$20.8 & 1.813 & 0.170 & \citet{koe98} \\
$^{*}$24.5 & 2.237 & 0.700 & \citet{koe98} \\
$^{*}$25.0 & 3.250 & 0.130 & IRAS \\
$^{*}$60.0 & 8.630 & 0.430 & IRAS \\
$^{*}$100.0 & 4.300 & 0.340 & IRAS \\
450.0 & 0.180 & 0.150 & \citet{grea99} \\
800.0 & $<$0.028 &  & \citet{jura95} \\
850.0 & 0.0191 & 0.0034 & \citet{grea99} \\
\hline \hline
\end{tabular}
\caption{\label{data}
Available infrared and submillimeter measurements.
Refer to section \ref{single} for the meaning of the mark ``$*$''.}
\end{center}
\end{table}
\subsection{Resolved mid infrared thermal data}
Mid-IR images revealed the morphology of the dust distribution
sensitive to these wavelengths.
Actually, the rather low resolution does not allow to
precisely describe the radial shape of the surface brightness along the major
axis and its vertical shape.
However, simple models are consistent with data if the suface density of
grains follows a radial power law $r^{-\Gamma}$ where $\Gamma$ is badly
constrained in the range $[0,2.5]$ (\citet{koe98} and \citet{jaya98}).
Assuming that the disk is optically and geometrically thin
(2-D disk model) and using a Bayesian approach, \citet{koe98} assessed a disk
inclination less than about $20\degr$ with respect to the line of sight.
Indeed, the vertical height to radial height ratio for a given isophote
at $20.8\,\mu$m is close to $0.35\pm 0.05$ which corresponds to a
flat disk inclined at $\sin^{-1}(0.35\pm 0.05) \simeq 20.5\pm 3\degr$ from
edge-on. Finally, both \citet{koe98} and \citet{jaya98} images are consistent
with a disk located at the position angle (PA) in the ranges~:
$[22\degr,34\degr]$ for \citet{koe98} and $[20\degr,40\degr]$ for
\citet{jaya98}.
\subsection{Resolved near infrared data}
\subsubsection{Reduction procedure}
The coronographic observations we reinvestigate in this paper have been first
presented in \citet{mou97}. This paper also describes the reduction procedure
used.
A main step in coronographic data reduction is the removal of the scattered
light remaining around the mask. This implies to use a comparison star
which scattered light is scaled to that of the object of interest. The scaling
factor is generally estimated by azimuthally averaging the division of the
star of interest by the comparison star. Azimuthal averaging over 360$\degr$
prevents from detecting pole on disks and decreases the chance of detection
of inclined disks or faint structures. Knowing the orientation of
the disk, we were able to estimate more precisely the
scaling factor in a section perpendicular to the disk.
\subsubsection{Radial excess almost in the direction of \hrb}
\label{Kimage}
\begin{figure}[tbph]
\begin{center}
\includegraphics[angle=0,origin=bl,width=0.49\textwidth]{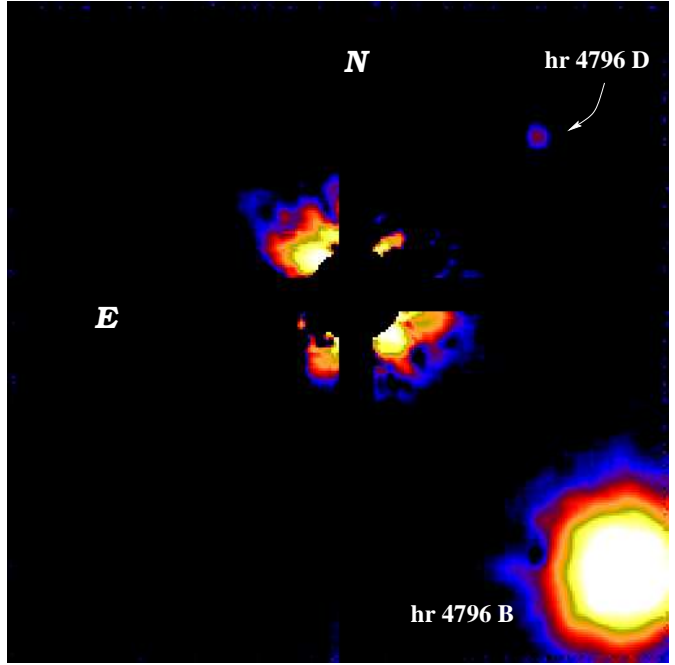}
\caption[]{Newly reduced coronographic image in logarithmic scale of \hra\
in K' filter revealing an excess close to the expected PA. The field-of-view
is 13''$\times$13''with a sampling of 0.05'' per pixel, smoothed to 0.2''.
The numerical mask occults up to about 60\,AU (0.9'') in radius. One can note
the faint companion \hrd\ (K'=14.5) at PA=311$\degr$ detected by \citet{kal93}
and \citet{mou97}.}
\label{hr4796_K}
\end{center}
\end{figure}

We present Figure \ref{hr4796_K} the newly reduced coronographic image
of \hra\ in K' band. An excess is present in the NE and SW regions.
Due to the spreading of the excess and due to the rather low signal to noise,
the position angle is badly constrained but is roughly consistent with both
the PA deduced from previous images and the direction of the physical
companion \hrb.
\begin{figure}[!h]
\begin{center}
\includegraphics[angle=90,origin=bl,width=0.50\textwidth]{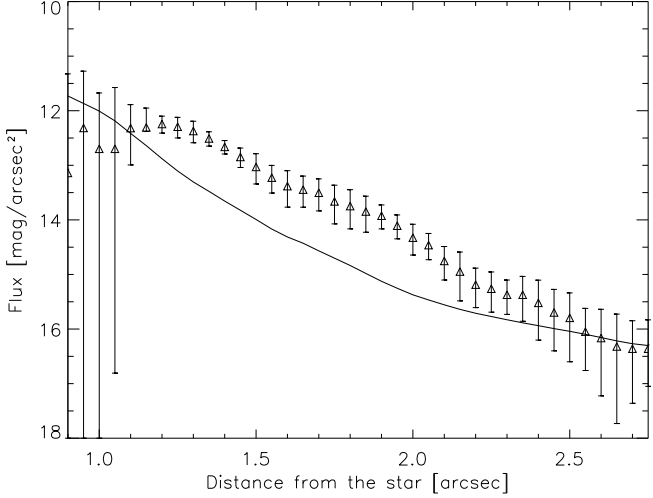}
\caption[]{
NE residual signal in the region around PA$=$30$\degr$ compared to
the noise (plain line). The latter is interpreted as the detection limit of the
disk. The residual signal in the regions perpendicular to the disk is close
to 0 on average (not represented in this logarithmic scale). On contrary,
the residual signal in the NE region is continuously greater
than the noise between 1.1''--1.2'' and 2.3''--2.4''.}
\label{limdetNE}
\end{center}
\end{figure}
\begin{figure}[!h]
\begin{center}
\includegraphics[angle=90,origin=bl,width=0.50\textwidth]{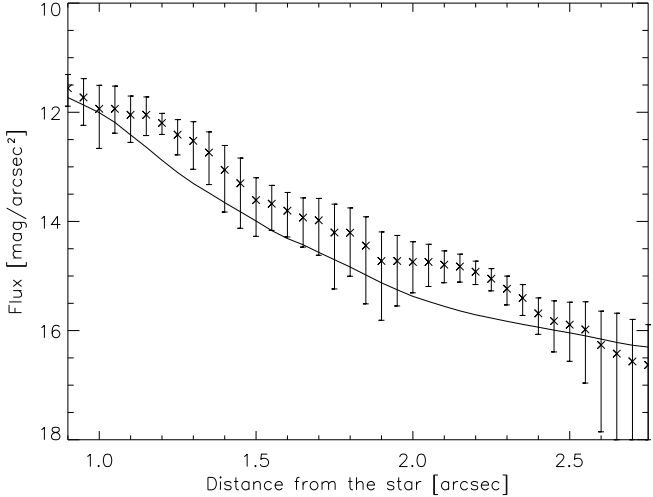}
\caption[]{Same as Figure \ref{limdetNE} but for the SW region
(around PA$=$210$\degr$).}
\label{limdetSW}
\end{center}
\end{figure}
We extract the residual signal azimuthally averaged in the ranges of PA~:
$[20\degr,40\degr]$ and $[200\degr,220\degr]$ as a function of
the distance from the star. We plot in Figures \ref{limdetNE} and
\ref{limdetSW} the comparison with the noise on the image estimated in the
regions perpendicular to the excess.
Between about 1.1'' and 2.3'', the NE and SW remaining signals
appear above the noise whereas in the perpendicular region the residual
signal fluctuates around 0 which characterizes the lack of \cs\ structure in
the NW and SE regions. Although the signal to noise ratio is
rather low close to the star, we find an azimuthally averaged surface
brightness at 1.1'' -- 1.2'' (75 -- 80\,AU) from the star~:
12.2$\pm0.5$\,mag.arcsec$^{-2}$ quite
similar to the range 11.7 -- 12.7\,mag.arcsec$^{-2}$ measured in H band by
\cite{sch99} between 1.1'' and 1.2''.
The observed surface brightness is roughly proportionnal to
$r^{-5}$ but deconvolved images would lead to steeper slopes as also
suggested by \citet{sch99} observations.
%
%
%
%
%
\section{Disk model assumptions}
\label{secmodel}
\subsection{Grain distribution}
We adopt a parametrical approach to describe the grain distribution in the 
disk similarly to that adopted to describe the \bp\ disk \citep{arty89}~:
\begin{eqnarray*}
n(r,z) = n_0\, R(r)\, Z(r,z)
\end{eqnarray*}
where  $n(r,z)$ is axisymmetrical grain density, $R(r)$ and $Z(r,z)$ are
respectively the radial and the vertical no-dimensional
distributions and $n_0$ the grain density at the normalized distance $r=r_0$
when $z=0$. The distance $r_0$ is fixed to 70\,AU in all the simulations and
$R(r)$ is normalized so as to have $R(r_0)=1$. 

Following \citet{arty89}, we assume a vertical distribution characterized by an
exponential with a shape depending on the parameter $\gamma$ ($\gamma>0)$~:
\begin{eqnarray*}
Z(r,z) = \exp\left(-\left(\frac{|z|}{\zeta(r)}\right)^{\gamma}\right)
\rm{\,with} \;\;\;\zeta(r)=\zeta_{r_0} \left(\frac{r}{r_0}\right)^{\beta}
\end{eqnarray*}
where the radial shape of the vertical scale height $\zeta(r)$ depends on
the parameter $\beta>0$. $\zeta_{r_0}$ is the normalized height when $r=r_0$.
This notation is useful because the integration of $Z(r,z)$ along the vertical
axis ($z$) of the disk is proportionnal to $r^{\beta}$. Therefore, the surface
density of the grains, which is the interesting quantity to model the
photometric data, can be written as~:
\begin{eqnarray*}
\sigma(r) = \int_{-\infty}^{+\infty} n(r,z)dz
= \sigma_0 \times R(r) \left(\frac{r}{r_0}\right)^{\beta}
\end{eqnarray*}
where $\sigma_0 = C_{\gamma} n_0 \zeta_{r_0}$ and
$C_{\gamma} = 2 \int_0^{+\infty} \exp(-x^{\gamma})dx$ ($x$ is an
non-dimensional variable).

We parametrize the radial distribution of the grains with a smooth
combination of two power laws~:
\begin{eqnarray*}
R(r) \propto \left\{\left(\frac{r}{r\dma{c}}\right)^{-2\alpha\dma{in}}
+ \left(\frac{r}{r\dma{c}}\right)^{-2\alpha\dma{out}} \right\}^{-\frac{1}{2}}\\
\end{eqnarray*}
with $\alpha\dma{in}>0$ and $\alpha\dma{out}<0$.
The inner disk is assumed to be cut-off at the distance corresponding to the
grain sublimation temperature ($\sim$ 1800\,K).
The maximum of the surface density $\sigma(r)$ is not $r\dma{c}$ but also
depends on $\alpha\dma{in}$, $\alpha\dma{out}$ and $\beta$~:
\begin{eqnarray}
r_{\mathrm{max}(\sigma)} = \left(\frac{\Gamma\dma{in}}
{-\Gamma\dma{out}}\right)^{\left(2\Gamma\dma{in}-
2\Gamma\dma{out}\right)^{-1}}\,\,r\dma{c}
\label{max2PL}
\end{eqnarray}
with $\Gamma\dma{in}=\alpha\dma{in}+\beta$ and
$\Gamma\dma{out}=\alpha\dma{out}+\beta$.
Outside this maximum, $\sigma(r)$ is roughly proportionnal to
$r^{\Gamma\dma{out}}$, and roughly proportionnal to $r^{\Gamma\dma{in}}$
inside.
Finally, the SED computation requires a four parameters grain
distribution~: $r\dma{c}$, $\Gamma\dma{in}$, $\Gamma\dma{out}$, and
$\sigma_0$.
\subsection{Grains characteristics}
\subsubsection{Composition assumptions~:}
We assume that the grains are porous aggregates made of a silicate core
coated by an organic refractory mantle (\citet{gre86} and references therein,
\citet{pol94}). Vaccum is assumed to fill the holes due to porosity.
If the grain temperature is less than the sublimation
temperature of the H$_2$O ice (T$\dma{sub,H_2O} \simeq 110-120$\,K),
the H$_2$O ice is able to fill all or part of the vaccum in the grain.
The relative abundances of each compound are given by the volume ratios~:
$q\dma{Sior}=\frac{V\dma{Si}}{V\dma{or}}$ and
$q\dma{SiorH_2O} = \frac{V\dma{Si}+V\dma{or}}{V\dma{H_2O}}$,
where $V\dma{Si}$, $V\dma{or}$ and $V\dma{H_2O}$ are respectively the
silicate, the organic refractories and the H$_2$O volumes.
The corresponding grain density $\rho\dma{g}$ is~:
\begin{eqnarray}
\rho\dma{g} = (1-P) * \left(\frac{q\dma{Sior}\,\rho\dma{Si}}{1+q\dma{Sior}}+
\frac{\rho\dma{or}}{1+q\dma{Sior}}+
\frac{\rho\dma{H_2O}}{q\dma{SiorH_2O}}\right)
\label{rhog}
\end{eqnarray}
where the individual densities are~: $\rho\dma{Si} = 3.5$\,g.cm$^{-3}$,
$\rho\dma{or} = 1.8$\,g.cm$^{-3}$ and $\rho\dma{H_2O} = 1.2$\,g.cm$^{-3}$.
Given the Si+or aggregate porosity $P$, the corresponding percentage of vaccum
removed due to the presence of the H$_2$O ice is therefore~:
\begin{eqnarray*}
p\dma{H_2O} = \left(\frac{1}{P}-1\right)\,\frac{100}{q\dma{SiorH_2O}}
\,\,\,\%\,\,\, .
\end{eqnarray*}

\citet{li98} proposed two extreme types of dust in Vega-like disks~:
amorphous or crystalline grains mainly depending on the evolution of the \cs\
environment  and probably depending on the spectral type.
In this paper, we investigate the two following types of grains~:
\\[-0.3cm]
\begin{enumerate}
\item \textit{``ISM-like grains''}~:
\begin{itemize}
\item amorphous Si and amorphous H$_2$O ice,
\item $q\dma{Sior} \simeq \frac{1}{2.12}$ for spherical grains and
$q\dma{Sior}\simeq \frac{1}{2.94}$ for cylindrical grains \citep{li97},
\item porosity $P$ between 0.45 \citep{mat96} and 0.8 \citep{mat89}.
\end{itemize}
\item \textit{``comet-like grains''}~:
\begin{itemize}
\item crystalline Si and crystalline H$_2$O ice,
\item $q\dma{Sior} \simeq \frac{1}{2}$ (comet Halley, \citet{gre98}),
\item porosity $P$ between 0.93 and 0.975 (comet Halley, \citet{gre90}).
\end{itemize}
\end{enumerate}
For all the models, we consider spherical grains and therefore we fixed the
volume ratio $q\dma{Sior}$ to $\frac{1}{2}$.
\begin{figure}[tbp]
\begin{center}
\includegraphics[angle=90,origin=bl,width=0.48\textwidth]{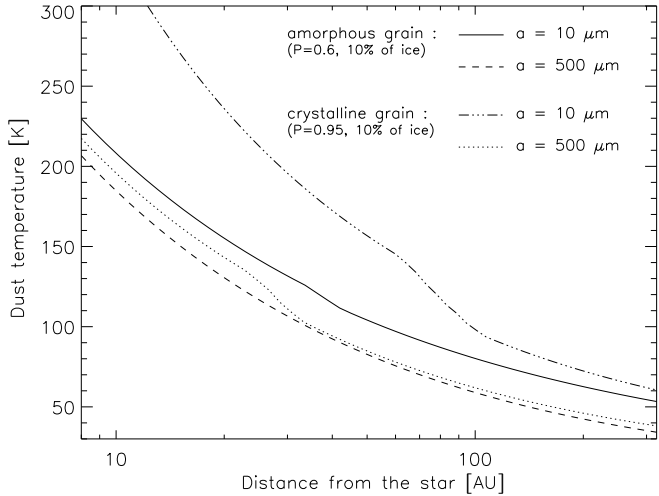}
\caption[]{Grain temperature versus the distance from the star for
amorphous and crystalline grains and for different grain sizes $a$.
The grain temperature has been computed from the energetic balance between
the absorbed flux and the emitted flux over all the spectrum.
We use a Kurucz spectrum for an A1V star scaled to the V
magnitude and extrapolated with a $\lambda^{-4}$ power law (Rayleigh-Jeans)
to simulate the star flux. The break in the grain temperature is due to
H$_2$O ice sublimation (see also Figure \ref{dSub120K}).
The dust temperature of the icy grains follows a radial
power law $r^q$ with $q$ between $-0.36$ and $-0.47$ for the amorphous grains
and $q$ between $-0.40$ and $-0.43$ for the crystalline grains.}
\label{temp}
\end{center}
\end{figure}
\subsubsection{Optical properties computation~:}
We compute
theorical absorption/emission and scattering efficiencies for spherical grains
via Mie theory \citep{boh83} knowing the ratio $x=\frac{a}{\lambda}$ where
$a$ is the grain radius and knowing $m(\lambda)$ the complex index of
refraction of the grain. In some extreme cases, Mie theory is not
efficient~: we used the Rayleigh-Gans theory when $|m(\lambda)|x>1000$ and 
$|m(\lambda)-1|x<0.001$ or the geometric optics when
$|m(\lambda)|x>1000$ and $|m(\lambda)-1|x>0.001$ \citep{laor93}.

We adopt the Maxwell-Garnett effective medium (MGEM) theory to compute
$m(\lambda)$ for the mixture (Si + ''or'' + H$_2$O ice and/or vaccum)
given the complex indexes of refraction of each compound, the volumes
ratios and the porosity.
We prefer the MGEM theory to the Bruggman effective medium (BEM) theory
because the MGEM theory considers that a mixture is composed of an inclusion
(the core) embedded in an homogeneous matrix (the mantle) whereas the
BEM theory treats both components in the same way \citep{boh83}.
\begin{figure}[tbp]
\begin{center}
\includegraphics[angle=90,origin=bl,width=0.48\textwidth]{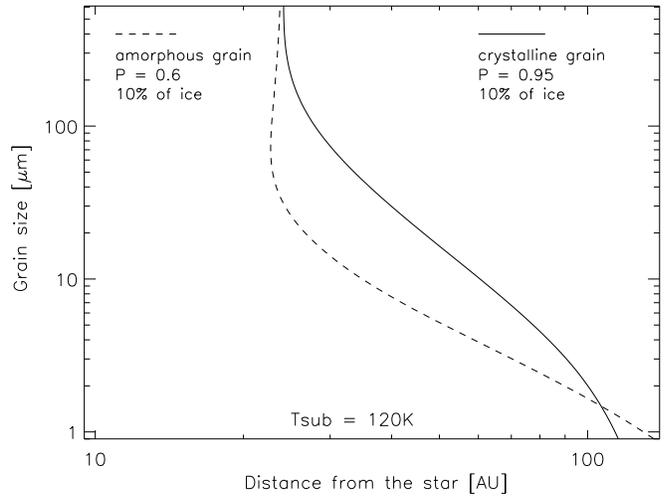}
\caption[]{Grain size corresponding to the H$_2$O ice sublimation
limit versus the distance from the star for a typical amorphous
grain and a typical crystalline grain. On the left side of the
curve, the ice is sublimated; on the right, grains are icy.}
\label{dSub120K}
\end{center}
\end{figure}
\subsubsection{Optical constants~:}
The wavelength dependence of the grain complex index of refraction $m(\lambda)$
depends on the individual compounds~:
\vspace*{-0.5cm}
\paragraph{$\rightarrow$ Silicates~:}
there is presently no consensus on what
could be ``representative'' astronomical silicates.
In the case of amorphous silicates, we arbitrarily adopt the modified
\citep{laor93} Draine \& Lee optical constants \citep{dlee94,drai95}.
In the case of crystalline silicates, we adopt the 
analytic complex indexes of refraction for crystalline olivine particles
proposed by \citet{muk90} between $\lambda=7\,\mu$m and $\lambda=200\,\mu$m
and we extrapolate (as \citet{li98}) with \citet{dlee94} optical
constants for $\lambda<7\,\mu$m and $\lambda>200\,\mu$m.
\vspace*{-0.5cm}
\paragraph{$\rightarrow$ Organic refractories~:}
we adopt the optical
constants computed by \citet{li97} (a full discussion on organic refractories
can be found in this paper).
\vspace*{-0.5cm}
\paragraph{$\rightarrow$ H$_2$O ice~:} in the case of amorphous ice and
between $\lambda=1.3\,\mu$m and $\lambda=144\,\mu$m, we use the optical
constants derived from \citet{fink} ($T=82$\,K) combined with the $100$\,K data
from \citet{hud93}.
In the case of crystalline ice and between $\lambda=0.89\,\mu$m and $\lambda=
333\,\mu$m, we adopt the hexagonal water ice optical constants
at a temperature of about $100$\,K derived by \citet{ock58} and \citet{ber69}.
For both the amorphous and crystalline ices, we extrapolate
with the \citet{war84} optical constants.
\subsubsection{Grain size distribution~:}
The choice of the grain size distribution $n(a)$ in the disk
is a key step in the grain modeling. We assume either a single grain
size ($a=a_{0}$) or a grain size distribution following the classical power
law $n(a)\ma{d}a \propto a^{-\kappa}\ma{d}a$ in the range
$[a\dma{min},a\dma{max}]$ which
may be better suited if the grains result from collisions among larger
bodies. We fixed ${\kappa}=3.5$, a typical value usually assumed for
collisional disks \citep{hell70,mat77}.
We use~:
$n(a)\ma{d}a = 2.5\,x^{-3.5}\ma{d}x \textrm{ \, with \, } x=a/a\dma{min}$
so as to have~:
\begin{eqnarray*}
\int_{1}^{x\dma{max}}2.5\,x^{-3.5}\ma{d}x  =1 \textrm{ \, with \, }
x\dma{max}=\frac{a\dma{max}}{a\dma{min}}\gg 1
\end{eqnarray*} 
\h\h\h
Finally, the grains are characterized by their amorphous or crystalline nature
and by the three free parameters~: $a_0$, $P$ and $p\dma{H_2O}$ for a single
grain size distribution or by the four free parameters~: $a\dma{min}$,
$a\dma{max}$, $P$ and $p\dma{H_2O}$ for a $-3.5$ power law grain size
distribution. Some examples of the dust temperature and of the grain size
corresponding to the H$_2$O ice sublimation limit versus the distance from the
star are plotted Figures \ref{temp} and \ref{dSub120K} respectively.
\begin{table*}[t] 
\begin{center}
\begin{tabular}[t]{c||ccc|ccc||c||cc|cc}
\hline \hline
\textbf{n$\uma{o}$} & $\Gamma\dma{in}$ & $\Gamma\dma{out}$ & $\begin{array}{c} \sigma_0 \\ \textrm{\scriptsize [grains.cm$^{-2}$]} \end{array}$ & $\begin{array}{c} a_0 \\ \textrm{\scriptsize [$\mu$m]} \end{array}$ & $P$ & $p\dma{H_2O}$ & \textbf{$\chi^2$} & $\begin{array}{c} a_0\uma{eff} \\ \textrm{\scriptsize [$\mu$m]} \end{array}$ & $P\dma{with H_2O}$ & $\begin{array}{c} \rho\dma{g} \\ \textrm{\scriptsize [g.cm$^{-3}$]} \end{array}$ & $\begin{array}{c} M\dma{d}(a_0) \\ \textrm{\scriptsize [$10^{-2}\,M_{\oplus}$]} \end{array}$ \\ 
\hline \hline
\noalign{\textit{amorphous}$\hspace{6cm}$}
\hline
\textbf{\#1} & 2.94 & -12.52 & 0.95 & 440 & 0.79 & 0.9 & \textbf{11.2} & 262 & 0.78 & 0.51 & 7.84 \\
\textbf{\#2} & 2.83 & -12.21 & 5.45 & 183 & 0.65 & 0.5 & \textbf{11.6} & 130 & 0.64 & 0.84 & 5.50 \\
\textbf{\#3} & 2.91 & -11.97 & 3.17 & 242 & 0.72 & 2.3 & \textbf{11.9} & 158 & 0.70 & 0.68 & 5.85 \\
\hline
\noalign{\textit{crystalline}$\hspace{6cm}$}
\hline
\textbf{\#4} & 2.88 & -10.49 & 8.11 & 147 & 0.93 & 38 & \textbf{5.7} & 59 & 0.58 & 0.59 & 3.04 \\
\textbf{\#5} & 3.11 & -7.27 & 1.06 & 391 & 0.975 & 19 & \textbf{6.7} & 114 & 0.79 & 0.28 & 4.21 \\
\textbf{\#6} & 3.30 & -11.85 & 0.74 & 505 & 0.975 & 15 & \textbf{7.5} & 148 & 0.83 & 0.23 & 4.09 \\
\hline \hline
\end{tabular}
\caption{\label{2PL70AU}
Sets of parameters providing the best fits of the 10 -- 100\,$\mu$m SED
assuming a  single grain size distribution and amorphous grains
(models \textbf{\#1}, \textbf{\#2} and \textbf{\#3}) or crystalline grains
(models \textbf{\#4}, \textbf{\#5} and \textbf{\#6}).
We also add some interesting quantitites such as~: the effective
grain size $a\uma{eff} = a\,(1-P)^{1/3}$ which represents the size of
a not porous grain with the same mass as the grain with the size $a_0$ and the
porosity $P$, the final grain porosity
$P\dma{with H_2O}=P\,(1-p\dma{H_2O}/100)$ taking into account that
$p\dma{H_2O}$\,\% of the vaccum has been removed and filled with H$_2$O ice,
the grain density $\rho\dma{g}$ (equation \ref{rhog})
and the disk mass $M\dma{d}(a_0)$.
For all the models, the peak of the surface density is close to
$r_{\mathrm{max}(\sigma)}=67\,$AU.}
\end{center}
\end{table*}
%
%
%
%
%
\section{Modelling}
\subsection{SED fitting assuming a single grain size distribution~:}
\label{single}
Our aim is to derive the dust distribution and to constrain the grains
properties with the available measurements.

We already have a constraint on the peak of the grain distribution
($r\dma{c}$). Indeed, the shape of the isophotes in scattered light at
$\lambda = 1.1\,\mu$m and $\lambda = 1.6\,\mu$mm \citep{sch99} suggests faint
ani\-so\-tro\-pic scattering properties for the grains at these wavelengths
(see also \ref{scatt}) implying that the observed maximum surface brightness
at roughly 70\,AU is probably close to the peak of the grain distribution.
We therefore fix $r\dma{c}=70$\,AU.

We let the 6 remaining parameters free in their respective ranges~:
\begin{itemize}
\item \textit{the 2 grains distribution parameters}~:\h
$\Gamma\dma{in}$ and $\Gamma\dma{out}$ in $[2$ , $13]$ and $[-13$ , $-2]$
respectively,
\item \textit{the 3 grains characteristics parameters}~:\h
grain size $a_0$ in $[1\,\mu$m , $600\,\mu$m$]$,
porosity $P$ in the range $[0.45$ , $0.8]$ if the grains are amorphous or
in $[0.93$ , $0.975]$ if the grains are crystalline and percentage
of H$_2$O ice $p\dma{H_2O}$ in $[0$\,\% , $50$\,\%$]$.
\item \textit{the normalized dust surface density $\sigma_0$}~:\h
given a set of the 5 previous parameters, the normalized dust surface density
$\sigma_0$ is used as a flux scaling factor so as to provide the best fit
of the SED.
\end{itemize}
\h
As a first step, we try to fit the 10 -- 100\,$\mu$m SED only.
Our aim is to minimize the $\chi^2$~:
\begin{eqnarray*}
\chi^2 = \sum_{i=1}^{N}\left( \frac{\Phi\dma{obs}(\lambda_{i}) -
\Phi\dma{sim}(\lambda_{i})}{\Delta\Phi\dma{obs}(\lambda_{i})} \right)^2
\end{eqnarray*}
where $\Phi\dma{sim}(\lambda_{i})$ are the simulated flux and $N$ the number
of data and discuss afterwards the consistency with thermal and
scattered light resolved images.

We start the minimization with a random vector of the parameters.
One has to note that we limit $N$ to the ten data marked with a ``$*$''.
Indeed, taking into account the numerous measurements at about
10\,$\mu$m, compared to the few data at larger wavelengths, would give a too
high weight to this SED region.
The $\chi^2$ probability $\wp(\chi^2,N-6)$ gives us a criterion for the
quality of the fit. For example, so as to have $\wp>10$\%
we need $\chi^2<7.78$ with $N=10$.
\begin{figure}[tbph]
\begin{center}
\includegraphics[angle=90,origin=bl,width=0.48\textwidth]{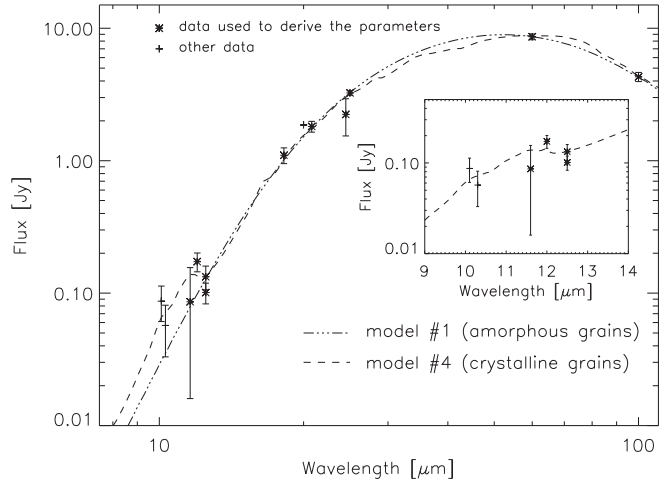}
\caption[]{Two fits of the 10\,$\mu$m -- 100\,$\mu$m Spectral Energy
Distribution of the \hra\ disk assuming a single grain size distribution
$a=a_0$ and a single grain distribution peaked at $r\dma{c}=70$\,AU from the
star (parameters from Table \ref{2PL70AU}).}
\label{set1et4}
\end{center}
\end{figure}
\begin{table*}[t] 
\begin{center}
\begin{tabular}[t]{c||c|cccc||c||ccc|ccc}
\hline \hline
\textbf{n$\uma{o}$} & $\begin{array}{c} \sigma_0 \\ \textrm{\scriptsize [grains.cm$^{-2}$]} \end{array}$ & $\begin{array}{c} a\dma{min} \\ \textrm{\scriptsize [$\mu$m]} \end{array}$ & $\begin{array}{c} a\dma{max} \\ \textrm{\scriptsize [cm]} \end{array}$ & $P$ & $p\dma{H_2O}$ & \textbf{$\chi^2$} & $\begin{array}{c} a\dma{min}\uma{eff} \\ \textrm{\scriptsize [$\mu$m]} \end{array}$ & $\begin{array}{c} a\dma{max}\uma{eff} \\ \textrm{\scriptsize [cm]} \end{array}$ & $P\dma{with H_2O}$ & $\begin{array}{c} \rho\dma{g} \\ \textrm{\scriptsize [g.cm$^{-3}$]} \end{array}$ & $\begin{array}{c} M\dma{d}(a\dma{min}) \\ \textrm{\scriptsize [$10^{-3}\,M_{\oplus}$]} \end{array}$ & $\begin{array}{c} M\dma{d} \\ \textrm{\scriptsize [$M_{\oplus}$]} \end{array}$ \\ 
\hline \hline
\noalign{\textit{amorphous}$\hspace{10cm}$}
\hline
\textbf{\#7} & 1.08$\times$10$^4$ & 4 & 18 & 0.57 & 39 & \textbf{18} & 3 & 14 & 0.35 & 1.29 & 3.78 & 1.59 \\
\textbf{\#8} & 8.10$\times$10$^3$ & 5 & 410 & 0.63 & 35 & \textbf{18} & 3 & 294 & 0.41 & 1.14 & 2.81 & 5.25 \\
\textbf{\#9} & 1.69$\times$10$^3$ & 10 & 137 & 0.8 & 12 & \textbf{27} & 6 & 80 & 0.70 & 0.59 & 3.15 & 2.31 \\
\hline
\noalign{\textit{crystalline}$\hspace{10cm}$}
\hline
\textbf{\#10} & 1.02$\times$10$^3$ & 12 & 246 & 0.93 & 18 & \textbf{54} & 5 & 101 & 0.77 & 0.36 & 1.98 & 1.78 \\
\textbf{\#11} & 1.27$\times$10$^2$ & 35 & 296 & 0.975 & 7 & \textbf{55} & 10 & 87 & 0.88 & 0.14 & 2.28 & 1.33 \\
\textbf{\#12} & 4.11$\times$10$^1$ & 64 & 116 & 0.975 & 4 & \textbf{84} & 18 & 34 & 0.94 & 0.10 & 3.15 & 0.85 \\
\hline \hline
\end{tabular}
\caption{\label{3.5FULLSED} Sets of parameters providing the best fits of the
full SED assuming a collisionnal grain size distribution ($\propto a^{-3.5}$)
and amorphous grains (models
\textbf{\#7}, \textbf{\#8} and \textbf{\#9}) or crystalline grains (models
\textbf{\#10}, \textbf{\#11} and \textbf{\#12}).
Refer to Table \ref{2PL70AU} for the definition of the quantities
$a\uma{eff}$, $P\dma{with H_2O}$ and $\rho\dma{g}$.
$M\dma{d}(a\dma{min})$ is the mass corresponding to the smallest grains
and $M\dma{d} = 2 M\dma{d}(a\dma{min}) \sqrt{a\dma{max}/a\dma{min}}$ is
the total disk mass. The peak of the surface density
$r_{\mathrm{max}(\sigma)}$ is close to $r\dma{c}$.}
\end{center}
\end{table*}
\begin{figure}[tbph]
\begin{center}
\includegraphics[angle=90,origin=bl,width=0.48\textwidth]{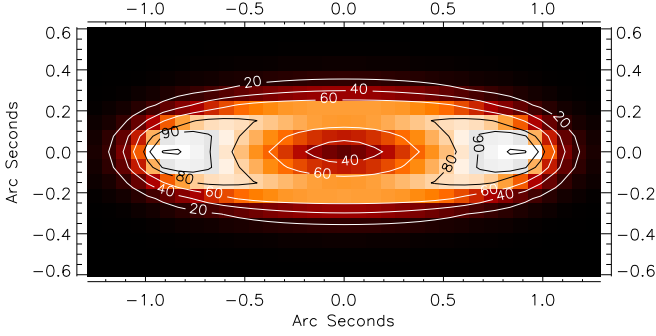}
\caption[]{Surface brightness contours of a 15$\degr$ tilted disk in
scattered light ($\lambda=1.1\,\mu$m) assuming the set of parameters
\textbf{\#4} (Table \ref{2PL70AU}) and isotropic scattering properties for the
grains. We also assume a flat disk~: $\beta=0.75$, $\zeta_{r_0}=4\ma{AU}$ and
an exponential vertical distribution~: $\gamma=1$.
The contour levels are in an arbitrary unit ; the upper level is
fixed to 100. In the inner part of the disk, the
surface brightness decreases slowly with the distance from the star ~: 
for instance, the surface brightness at 0.5'' from the distance corresponding
to the upper level still represents 60\% of the maximum surface brightness.
Such an inner dust distribution ($\Gamma\dma{in}\simeq 3$) is
then not consistent with HST observations \citep{sch99}.}
\label{contours}
\end{center}
\end{figure}
We present Table \ref{2PL70AU} six different sets of the parameters which
lead to the best 10 -- 100\,$\mu$m SED fitting assuming a single grain size
distribution $a_0$ and a single grain distribution.

The 3 crystalline grains models satisfy the $\chi^2$ criterion~: 
$\wp>10$\,\% whereas the amorphous grains models do not.
Amorphous grains do not provide a proper fit of the 10 -- 12.5\,$\mu$m data.
The crystalline hypothesis reproduce very well the 10\,$\mu$m emission
feature. Furthermore,
the 10.1\,$\mu$m and 10.3\,$\mu$m data, not used to derive the parameters,
are also well fitted with the model (Figure \ref{set1et4}).
Note that the 20\,$\mu$m -- 100\,$\mu$m continuum is fitted both
with crystalline grains and with amorphous grains.

Grains with $a_0\uma{eff}>60\,\mu$m are needed to reproduce
the shape of the SED (see Table \ref{2PL70AU} for the
definition of $a_0\uma{eff}$) which leads to grains sizes $a_0$ larger than
$150\,\mu$m.
Most of the grains are located further than the H$_2$O ice
sublimation distance (Figure \ref{dSub120K}) and have to be significantly icy~:
$p\dma{H_2O}>15-20$\,\% (consistent with \citet{jura98} assumptions).

We find $\Gamma\dma{in}$ to be close to 3 and the outer surface density
($\Gamma\dma{out}$) to be steeper than $r^{-7}$ (more probably as steep as
$r^{-10}$). Whereas the outer slope of the grain surface density is consistent
with the scattered ligth images, the inner slope is not. It would induce much
more light in the inner disk than observed.
Indeed, an $r^{3}$ inner surface density yields an inner midplane surface
brightness in scattered light less steep than at least $r^{2-\beta}$ (case of
an edge-on disk assuming isotropic scattering properties for the grains,
\citet{nak90}). As an example, we plot Figure \ref{contours} the surface
brigthness contours for a 15$\degr$ tilted (from edge-on) disk in scattered
light at $\lambda=1.1\,\mu$m.

\begin{figure}[tbph]
\begin{center}
\includegraphics[angle=90,origin=bl,width=0.48\textwidth]{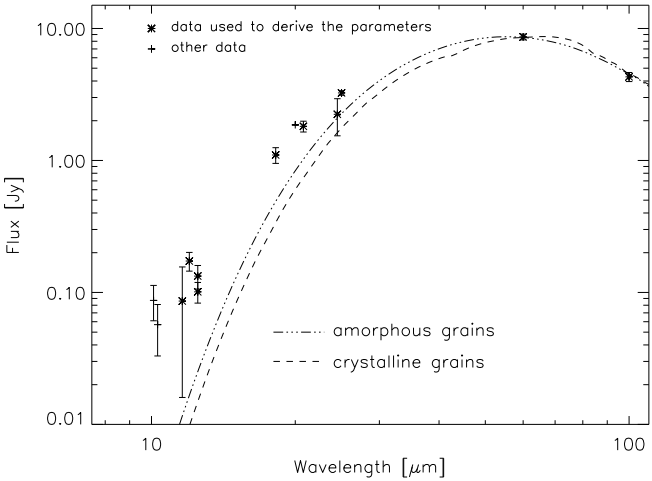}
\caption[]{Same as Figure \ref{set1et4} but assuming $\Gamma\dma{in}=10$.}
\label{Gammain10}
\end{center}
\end{figure}

Taking into account the constraint from scattered light images, we therefore
try to derive the grains properties which could reproduce the 10 --
100$\,\mu$m SED with $\Gamma\dma{in}\simeq 10$.
Such an inner slope acts
as a cut-off for the short wavelengths efficient emitters which are located
close to the star. Under these assumptions neither the crystalline nor the
amorphous grains provide good fits of the 10 -- 100$\,\mu$m SED (Figure
\ref{Gammain10}).
In the present case, we also exclude to add an inner dust distribution
(typically inside 50\,AU) to fit the SED at short wavelengths.
In this scheme, the inner and outer disks would have similar contributions
(flux) at 20\,$\mu$m (Figure \ref{Gammain10}), which is not compatible with
20\,$\mu$m resolved images \citep{koe98,jaya98}.

In conclusion, assuming a single grain size distribution, we are able to
properly fit the 10 -- 100\,$\mu$m SED but not the resolved images.
Moreover, the presence of hot crystalline grains seems necessary to
fit the 10\,$\mu$m measurements.
%
%
%
%
%
\subsection{SED fitting assuming a grain size distribution proportionnal to
$a^{-3.5}$~:}
\label{double}
We now assume a collisional grain size distribution $n(a)$ proportionnal to
$a^{-3.5}$ between a minimum and a maximum grain size $a\dma{min}$ and
$a\dma{max}$. We also assume that the relative abundances of the different
compounds and the porosity do not depend on the grain size.

We adopt a spatial grain distribution consistent with the scattered light
images and therefore fix $r\dma{c}$ to 70\,AU, the inner slope
$\Gamma\dma{in}$ to $10$ and the outer slope $\Gamma\dma{out}$ to $-11.5$.
The grain properties are now the free parameters left.
\subsubsection{Single grain population~:}
\label{asingle}
We try to fit the SED by minimizing the $\chi^2$ associated to
the 10 data marked with a ``$*$'' in Table \ref{data} plus the 850\,$\mu$m
measurement.
We keep the porosity $P$ and the percentage of H$_2$O ice free in the same
ranges as for the single grain size distribution and investigate the
respective broad ranges $[1\,\mu$m , $500\,\mu$m$]$ and $[600\,\mu$m ,
$5\,$m$]$ for the minimum $a\dma{min}$ and the maximum $a\dma{max}$ grain
sizes.

\begin{figure}[tbph]
\begin{center}
\includegraphics[angle=90,origin=bl,width=0.48\textwidth]{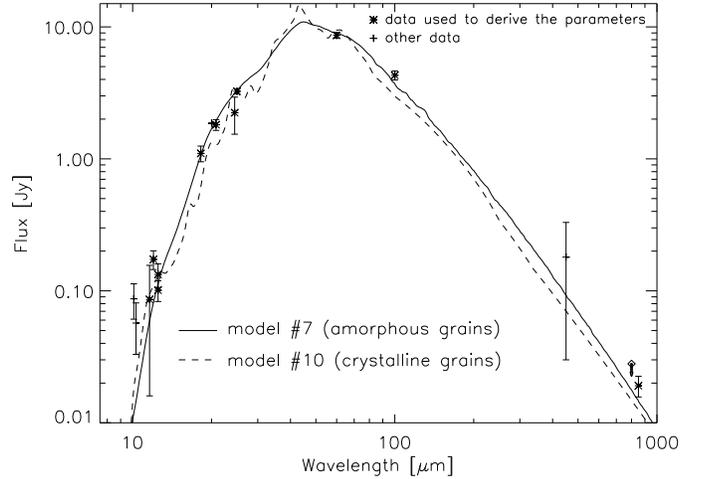}
\caption[]{Two fits of the full SED assuming a collisional grain size
distribution and a single dust population peaked at 70\,AU from the star.
The 10\,$\mu$m and 850\,$\mu$m data are not properly fitted.}
\label{set7et10}
\end{center}
\end{figure}

The results are summarized in Table \ref{3.5FULLSED}.
Whatever the nature of the grains, the fits do not satisfy the
$\chi^2$ criterion ($\wp>10$\% if $\chi^2<9.24$),
but micronic to centimetric amorphous grains better reproduce the
shape of the SED continuum than crystalline grains (Figure \ref{set7et10}).
Therefore, crystalline grains (if any) probably represent a minor part of the
grain population in the annulus peaked at 70\,AU.
On the other hand, the 10\,$\mu$m measurements are not well reproduced with the
amorphous grain model still suggesting (see previous section) the presence of a
small amount of crystalline grains in order to significantly emit at 10\,$\mu$m
but not at larger wavelengths.
If there are crystalline grains at 70\,AU, they are neither numerous
nor hot enough to produce such a spectral energy distribution.
Therefore, if the 10\,$\mu$m data are not overestimated (especially the
10.1\,$\mu$m, the 10.3\,$\mu$m and the 12\,$\mu$m measurements), we can
exclude to fit the full SED from the NIR to the submillimeter wavelengths with
the same dust population.
\begin{table*}[t] 
\begin{center}
\begin{tabular}[t]{c||c|ccc||c||ccc|ccc}
\hline \hline
\textbf{n$\uma{o}$} & $\begin{array}{c} \sigma_0 \\ \textrm{\scriptsize [grains.cm$^{-2}$]} \end{array}$ & $\begin{array}{c} a\dma{min} \\ \textrm{\scriptsize [$\mu$m]} \end{array}$ & $\begin{array}{c} a\dma{max} \\ \textrm{\scriptsize [cm]} \end{array}$ & $p\dma{H_2O}$ & \textbf{$\chi^2$} & $\begin{array}{c} a\dma{min}\uma{eff} \\ \textrm{\scriptsize [$\mu$m]} \end{array}$ & $\begin{array}{c} a\dma{max}\uma{eff} \\ \textrm{\scriptsize [cm]} \end{array}$ & $P\dma{with H_2O}$ & $\begin{array}{c} \rho\dma{g} \\ \textrm{\scriptsize [g.cm$^{-3}$]} \end{array}$ & $\begin{array}{c} M\dma{d}(a\dma{min}) \\ \textrm{\scriptsize [$10^{-3}\,M_{\oplus}$]} \end{array}$ & $\begin{array}{c} M\dma{d} \\ \textrm{\scriptsize [$M_{\oplus}$]} \end{array}$ \\ 
\hline \hline
\noalign{\textit{amorphous}$\hspace{10cm}$}
\hline
\textbf{\#13} & 1.79$\times$10$^3$ & 10 & 143 & 3.03 & \textbf{1.38} & 7 & 106 & 0.58 & 0.97 & 5.10  & 3.87 \\
\textbf{\#14} & 1.65$\times$10$^3$ & 10 & 291 & 2.12 & \textbf{2.06} & 7 & 214 & 0.59 & 0.96 & 5.10 & 5.43 \\
\textbf{\#15} & 1.38$\times$10$^3$ & 11 & 36 & 0.13 & \textbf{2.72} & 8 & 27 & 0.60 & 0.95 & 5.58 & 2.01 \\
\hline
\noalign{\textit{crystalline}$\hspace{10cm}$}
\hline
\textbf{\#16} & 2.70$\times$10$^1$ & 80 & 208 & 2.95 & \textbf{17.3} & 33 & 86 & 0.90 & 0.20 & 8.00 & 2.58 \\
\textbf{\#17} & 1.12$\times$10$^2$ & 38 & 359 & 6.54 & \textbf{22.8} & 16 & 148 & 0.87 & 0.24 & 4,53 & 2.76 \\
\textbf{\#18} & 9.94$\times$10$^1$ & 41 & 276 & 6.49 & \textbf{23.5} & 17 & 114 & 0.87 & 0.24 & 4.80 & 2.50 \\
\hline \hline
\end{tabular}
\caption{\label{3.5OUTERSED} Sets of parameters providing the best fits of the
20 -- 850\,$\mu$m SED assuming a collisionnal grain size distribution
($\propto a^{-3.5}$) and
amorphous grains (models \textbf{\#13}, \textbf{\#14}
and \textbf{\#15}) or crystalline grains (models \textbf{\#16}, \textbf{\#17}
and \textbf{\#18}).
Refer to Table \ref{3.5FULLSED} for the definition of the quantities
$a\uma{eff}$, $P\dma{with H_2O}$, $\rho\dma{g}$, $M\dma{d}(a\dma{min})$
and $M\dma{d}$.
The peak of the surface density $r_{\mathrm{max}(\sigma)}$ is close to
$r\dma{c}$.}
\end{center}
\end{table*}
\subsubsection{Two grain populations~:}
\label{adouble}
\paragraph{$\rightarrow$ The cold annulus~:}\h
\label{cold}
We now assume that the excesses at large wavelengths mainly come from
the annulus resolved in scattered light.
This is consistent in terms of spatial extension with the disk observed
at 20.8\,$\mu$m \citep{koe98} suggesting that the same dust population
may be responsible for both the 20.8\,$\mu$m and the 1.1\,$\mu$m images.

We therefore try to fit the 5 data at wavelengths larger than 20\,$\mu$m
(excluding the 24.5\,$\mu$m and 450\,$\mu$m data due to too high uncertainties)
and try to constrain the grains properties consistent with the colder part
of the SED.
We fix the porosity $P$ to 0.6 for amorphous grains or to 0.93 for
crystalline grains as suggested by the full SED fitting (Table \ref{3.5FULLSED}).
We keep the three parameters $a\dma{min}$, $a\dma{max}$ and $p\dma{H_2O}$
free in the same ranges as in the previous subsection (\ref{asingle}).

\begin{figure}[tbph]
\begin{center}
\includegraphics[angle=90,origin=bl,width=0.48\textwidth]{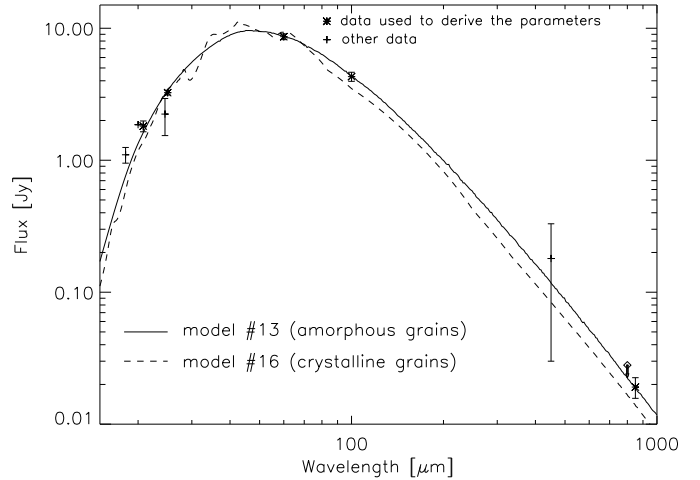}
\caption[]{Two fits of the 20 -- 850\,$\mu$m SED assuming the dust resolved in
scattered light and peaked at 70\,AU from the star is responsible for the excesses
at wavelengths larger than about 20\,$\mu$m.}
\label{set13et16}
\end{center}
\end{figure}

The results of the $\chi^2$ minimization are summarized in Table
\ref{3.5OUTERSED} and Figure \ref{set13et16} shows the best fits for
each grain nature.
The cold SED continuum is remarkably well fitted ($\wp > 10$\% if
$\chi^2<2.71$) with amorphous grains whereas the crystalline grains lead to
very poor $\chi^2$.
Very little icy grains ($p\dma{H_2O}<5$\%) with sizes
larger than 10\,$\mu$m (15\,$\mu$m if we assume $P=0.7$) are required to fit
the 20 -- 25\,$\mu$m data and bodies as large as a few meters are necessary to
correctly reproduce the submillimeter excess.
This leads to an outer disk mass larger than a few Earth masses.
\paragraph{$\rightarrow$ The hot dust population~:}\h
At 20.8\,$\mu$m, the simulated cold annulus (model \textbf{\#13}) contributes
to about 90\% of the measured excess at this wavelength. Therefore and
conversely to
the single grain size model, the addition of a second grain population would
not modify the consistency between the simulated thermal images at
20.8\,$\mu$m and the observations.

\begin{figure*}[th]
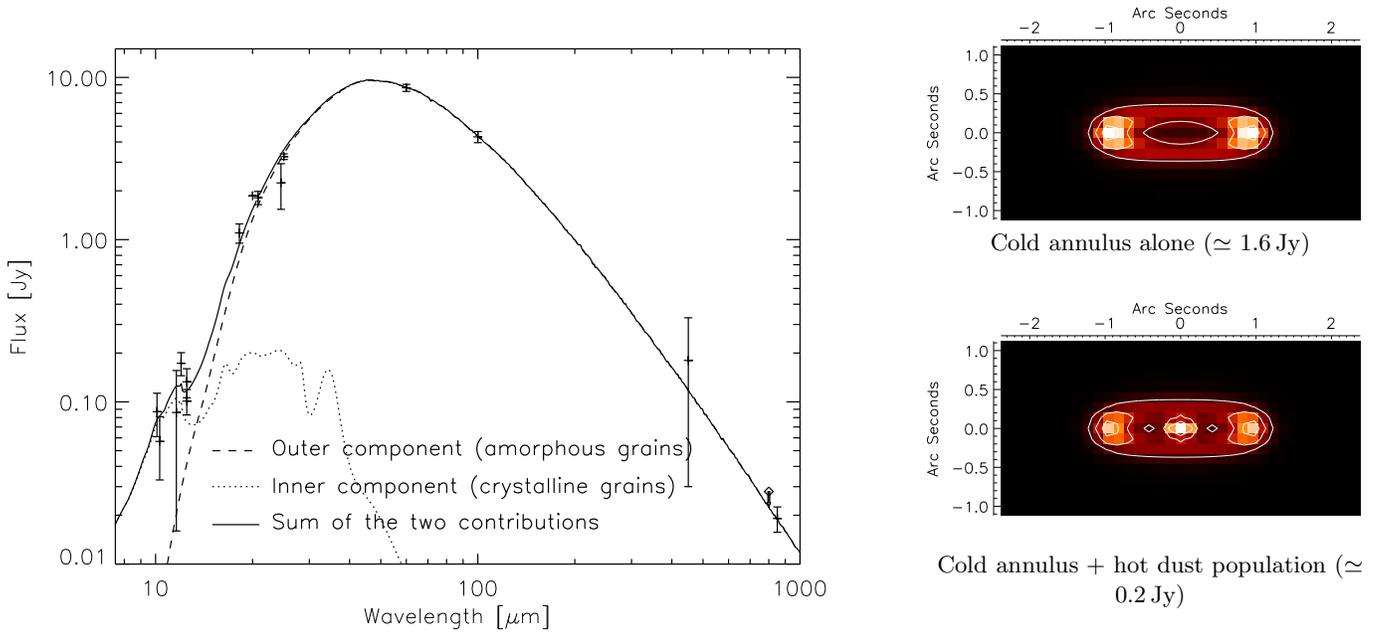

\hbox to \textwidth
{
\parbox{0.6\textwidth}{
\includegraphics[angle=90,width=0.59\textwidth,origin=br]{8413.f11}}
\hfil
\parbox{0.33\textwidth}{
\parbox{0.33\textwidth}{
\includegraphics[angle=90,width=0.32\textwidth,origin=br]{8413.f12}}
\vspace{1truemm}\par\noindent
\parbox{0.33\textwidth}{\begin{center} Cold annulus alone ($\simeq$ 1.6\,Jy)
\end{center}}
\vspace{1truemm}\par\noindent
\parbox{0.33\textwidth}{
\includegraphics[angle=90,width=0.32\textwidth,origin=br]{8413.f13}}
\vspace{1truemm}\par\noindent
\parbox{0.33\textwidth}{\begin{center} Cold annulus + hot dust population
($\simeq$ 0.2\,Jy) \end{center}}
}}
\caption[]{\textit{Left}~: the full SED fitting assuming two dust
populations~: model \textbf{\#13} $+$ an inner dust population (see text).
\textit{Right}~: simulated disks in thermal
emission at $\lambda = 20.8\,\mu$m assuming grain properties and surface
density derived from the SED fitting. We also assume a vertical structure
quite similar to the \bp\ disk one. We simulate the disks with the same pixel
size (0.14'') as the observations performed by \citet{koe98}. The images have
been convolved with an Airy pattern to take into account the PSF of a 10\,m
telescope at $\lambda = 20.8\,\mu$m.}
\label{2pop}
\end{figure*}

We therefore subtract the simulated flux coming from the outer annulus (model
\textbf{\#13}) to the observed excesses and try to fit the remaining excesses
with a second dust distribution made of crystalline grains (suggested by the
single grain size distribution fitting (section \ref{single})).
The full SED assuming two grain populations is now very well fitted (see
Figure \ref{2pop}). In this example, the inner dust distribution represents
a mass of about $3.5\times 10^{-5}\,M_{\oplus}$ made of 450\,$\mu$m
crystalline grains ($P=0.97$) peaked at 9\,AU from the star and heated at
about 210 -- 220\,K (see for example Figure \ref{temp}).
%
%
%
%
%
%
\begin{figure*}[tbph]
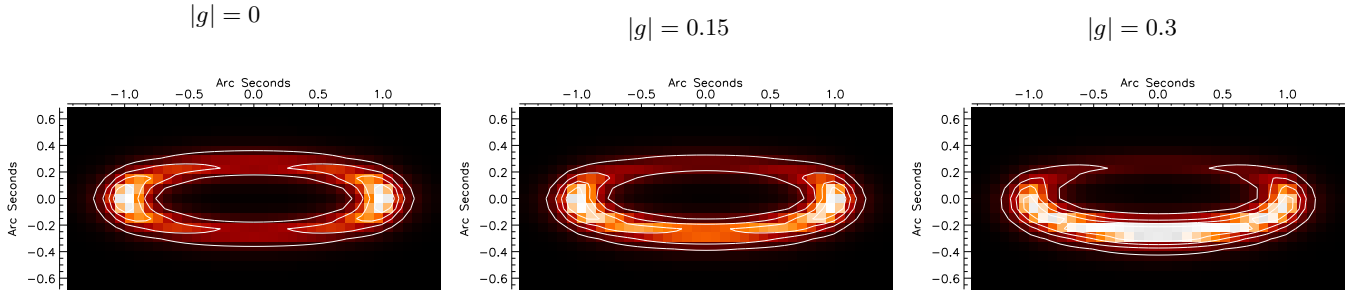

\hbox to \textwidth
{\parbox{0.33\textwidth}{\begin{center} $|g| = 0$
\end{center}}
\hfil
\parbox{0.33\textwidth}{\begin{center} $|g| = 0.15$
\end{center}}
\hfil
\parbox{0.33\textwidth}{\begin{center} $|g| = 0.3$
\end{center}}}
\vspace{1truemm}\par\noindent
\hbox to \textwidth
{
\parbox{0.33\textwidth}{
\includegraphics[angle=90,width=0.32\textwidth,origin=br]{8413.f14}}
\hfil
\parbox{0.33\textwidth}{
\includegraphics[angle=90,width=0.32\textwidth,origin=br]{8413.f15}}
\hfil
\parbox{0.33\textwidth}{
\includegraphics[angle=90,width=0.32\textwidth,origin=br]{8413.f16}}}
\caption[]{Simulations of the cold annulus peaked at 70\,AU in scattered light
at $\lambda = 1.1\,\mu$m for
different asymmetry factor $|g|$ ; the inner hot dust not observable has not
been added. We assume the same spatial distribution and grains properties
parameters as Figure \ref{2pop}. The pixel size is 0.076'' \citep{sch99}.
}
\label{scbpic}
\end{figure*}
\subsection{Images}
\label{images}
We now adopt the two grain populations model from previous section
(\ref{adouble})~: an amorphous dust population peaked at 70\,AU (model
\textbf{\#13}) plus a crystalline dust population at 9\,AU from the star.
We assume a disk inclination $i_{\mathrm{tilt}}$ of 15$\degr$ with
respect to the line of sight.
\subsubsection{Thermal images}
We simulate the disk in thermal emission so as to
derive further informations
on the vertical shape of the disk (parameters $\zeta_{r_0}$, $\gamma$ and
$\beta$ assumed to be constant over all the disk).
We actually have very few constraints on the vertical shape of the disk, also
we arbitrarily start with parameters derived for \bp\ disk~:
opening angle of the disk~$\sim 6\degr$ (\textit{i.e.} $\zeta_{r_0} \sim
4$\,AU), $\beta \sim 0.75$ and $\gamma \sim 1$ \citep{mou97}.
We plot (Figure \ref{2pop}) the simulated disk at $\lambda =
20.8\,\mu$m. The simulated vertical height to radial height ratios for a
given isophote are between 0.32 and 0.35~; they compare very well with the
values 0.35$\pm$0.5 from \citet{koe98}.

If we afterwards simulate the thermal images for different sets of the
parameters $\zeta_{r_0}$, $\beta$ and $\gamma$, we get a few interesting
results~:
\begin{itemize}
\item The normalized height $\zeta_{r_0}$ at $r_0=70$\,AU is probably less than
about 7 -- 8\,AU leading to opening angles of the disk smaller than about
15$\degr$. For example, $\zeta_{r_0}=10$\,AU provides a too large vertical
structure and the vertical spreading of the dust yields a smaller surface flux
density for the outer annulus compared to the inner one. Such a high constrast
is not observed.
\item The parameter $\beta$ is badly constrained~: 0.75 provides a good shape
of the isophotes (Figure \ref{2pop}) but 1.5 might also be a correct value.
\item Given the resolution of the observations, the parameter $\gamma$ is
completly undetermined.
\end{itemize}
\subsubsection{Scattered light images}
\label{scatt}
To simulate the scattered light images, we adopt
the empirical Henyey-Greenstein phase function \citep{HG}
which depends on the single parameter $|g|$ ($0$ for isotropic scattering
and $1$ for total forward or backward scattering). Indeed, theoretical phase
functions from Mie theory are probably not realistic for porous fluffy grains.

We simulate at $\lambda=1.1\,\mu$m the \textit{cold annulus}, which is
responsible for the scattered light images, assuming different values of the
parameter $|g|$.
As anticipated, due to the disk inclination with respect to the line of sight,
the morphology of the observed disk strongly depends on the anisotropic
scattering properties (Figure \ref{scbpic}).
The observed isophotes at $\lambda = 1.1\,\mu$m \citep{sch99} implies that the
asymmetry factor $|g|$ is smaller than 0.15.

At 1.1\,$\mu$m, the simulated cold annulus scattered a total flux close
to 8.6\,mJy (5.0\,mJy at 1.6\,$\mu$m) assuming $|g|\simeq 0$.
The integrated flux density corresponding to the detected part of the annulus
with HST (outside 0.65'' in radius) represents 5.2\,mJy at  $\lambda =
1.1\,\mu$m, \textit{i.e.} only slightly smaller than 7.5$\pm$0.5\,mJy from
\citet{sch99}.

We also simulate the disk in K' band. At 1.1'' from the star along the major
axis of the annulus, the flux density
is about 16\,$\mu$Jy in a 0.05''$\times$0.05'' pixel. The corresponding
surface brightness is about 12.5\,mag.arcsec$^{-2}$ consistent with the
12.2$\pm$0.5\,mag.arcsec$^{-2}$ measured between 1.1'' and 1.2'' (subsection
\ref{Kimage}).
%
%
%
%
%
\subsection{Summary of model results}
Our model involves two dust components~:
\begin{itemize}
\item a cold annulus peaked at 70\,AU from the star made of ISM-like grains
in terms of chemical composition (amorphous) and porosity ($P\sim0.6$) and a
very few icy. Assuming a collisional grain size distribution ($\propto
a^{-3.5}$), we infer a minimum grain size in the ring of $10\,\mu$m and we
find that bodies as large as at least a few
meters may have already been formed. This leads to a total dust mass of a few
Earth masses held in a very narrow annulus.
\item an inner dust population at about 9 -- 10\,AU from the star made of very
porous ($P\sim0.97$) crystalline grains responsible for the emission feature at
$\lambda\sim10\,\mu$m and for the slight excess centered on the star in the
$20.8\,\mu$m images \citep{koe98}.
\end{itemize}
%
%
%
\section{Implications of the results on the disk dynamics}
\label{discuss}
\subsection{Gas to Dust ratio}
\label{gdratio}
It is of interest to assess the gas to dust ratio in the disk.
Given the upper limit on the gas mass $M\dma{g} < 1 - 7\,M_{\oplus}$
derived by \citet{grea99} and the dust mass $M\dma{d} \simeq 4\,M_{\oplus}$
from model \textbf{\#13} which can be reasonably considered as a lower limit
(see also \ref{Mloss}),
we infer a gas to dust ratio less than 1. In this scheme, the gas would
probably not play anylonger an important role, as claimed by \citet{grea99}.
\subsection{Radiation pressure}
\begin{figure}[!h]
\begin{center}
\includegraphics[angle=90,origin=bl,width=0.50\textwidth]{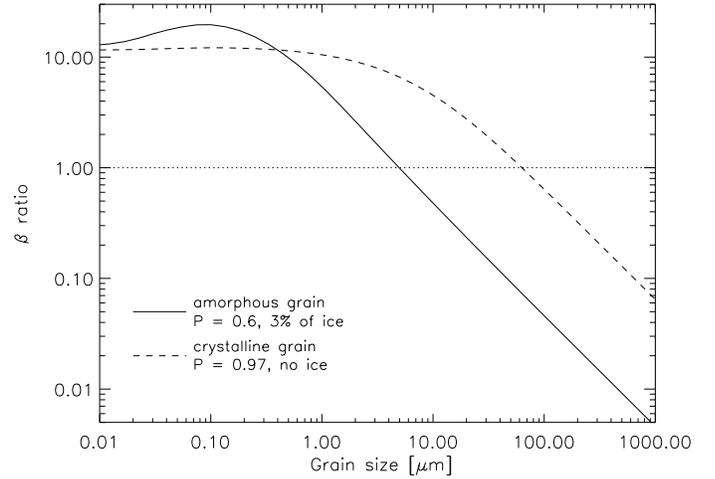}
\caption[]{Radiation pressure force to the gravitationnal force ratio
($\beta\dma{pr}$) assuming $L_{*}=20\,L_{\odot}$ and $M_{*}=2.5\,M_{\odot}$. 
The amorphous grains characterize the cold dust annulus (solid ligne)
and the crystalline ones the hot dust population (dashed ligne).
Below the dotted ligne, the grains are gravitationally bound.}
\label{beta}
\end{center}
\end{figure}
The ratio $\beta\dma{pr}$ of radiation pressure to gravity as a function of
the grain size is given by~:
\begin{eqnarray*}
\beta\dma{pr} = \frac{3L_{*}\langle Q\dma{pr}(a)\rangle}{16\pi G M_{*}c a
\rho\dma{g}}
\textrm{ \, with \, } \langle Q\dma{pr}(a)\rangle = \frac{\int Q\dma{pr}(a)
F_{*}(\lambda)d\lambda} {\int F_{*}(\lambda)d\lambda}\,\, .
\end{eqnarray*}
assuming $L_{*}=20\,L_{\odot}$ and $M_{*}=2.5\,M_{\odot}$.
We find a blow-out grain size limit ($\beta\dma{pr}=1$) close to
$5\,\mu$m for the amorphous grains in the cold annulus (Figure \ref{beta}).
Also, amorphous grains with sizes between $10\,\mu$m ($\beta\dma{pr}\simeq
0.5$) and about $45\,\mu$m ($\beta\dma{pr}\simeq 0.1$) have orbits
significantly excentric.
Interestingly, the smallest grains we find in our fits in the cold annulus~:
$a\dma{min}\simeq 10\,\mu$m is very close to the blow-out size limit.
For the crystalline grains at 9\,AU from the star, the blow-out
grain size limit is close to $65\,\mu$m (Figure \ref{beta}).
The grains deduced from our modelling are then gravitationally bound.
\subsection{Times scales and dominant processes}
The life times of the grains in the disk under Poynting-Robertson drag,
collisions and sublimation, can be written respectively as
(e.g. Backman \&\ Paresce 1993)~:
\begin{eqnarray*}
t\dma{PR} = \frac{4 \pi c^2}{3} \frac{a \rho\dma{g}r^2}
{\langle Q\dma{abs}\rangle L_{*}}\simeq 35 \frac{\rho\dma{g}}
{\langle Q\dma{abs}\rangle}
a_{\mu\mathrm{m}} r_{\mathrm{AU}}^2 \,\textrm{\,\,yr}
\end{eqnarray*}
where $\langle Q\dma{abs}\rangle$ is the absorption efficiency
averaged over the star spectrum,
\begin{eqnarray*}
t\dma{coll} \simeq \frac{1}{2\pi \langle a^2\rangle\sigma(r)\Omega}
\simeq \frac{2.5\times 10^{6}r_{\mathrm{AU}}^{\frac{3}{2}}}
{\langle a_{\mu\mathrm{m}}^2\rangle\,\,\sigma(r)}\sqrt{\frac{M_{\odot}}{M_{*}}}
\textrm{\,\,\,yr}
\end{eqnarray*}
where $\Omega=\sqrt{GM_*}/r^{3/2}$
is the Keplerian circular rotation frequency at radius $r$, and
$\langle a^2\rangle$ the grain size square averaged over the grain size
distribution,
\begin{eqnarray*}
t\dma{sub} = \frac{a\dma{eq} \rho\dma{H_2O}}{6.7\times10^{11}}
\frac{10^{\frac{2480}{T\dma{g}}}}{T\dma{g}^{3.5}} 
\simeq 2\times 10^{-16}a_{\mathrm{eq},\mu\mathrm{m}}
\frac{10^{\frac{2480}{T\dma{g}}}}{T\dma{g}^{3.5}} \textrm{\,\,\,yr}
\end{eqnarray*}
where $a\dma{eq} = a\,
\left(P\,p\dma{H_2O}/100\right)^{1/3}$ is the equivalent grain radius for a
pure ice sphere with the same mass of ice as in a grain with a size $a$.
The time-scales for the cold annulus and the hot dust population are
summarized in Table \ref{times}.
\begin{table}[!h] 
\begin{center}
\begin{tabular}[h]{cc|ccc}
\hline \hline
$\begin{array}{c} \textrm{Dust} \\ \textrm{population} \end{array}$ & $\begin{array}{c} r \\ \textrm{\scriptsize $[$AU$]$} \end{array}$ & $\begin{array}{c} t\dma{PR} \\ \textrm{\scriptsize $[$yr$]$} \end{array}$ & $\begin{array}{c} t\dma{coll} \\ \textrm{\scriptsize $[$yr$]$} \end{array}$ & $\begin{array}{c} t\dma{sub} \\ \textrm{\scriptsize $[$yr$]$} \end{array}$ \\
\hline \hline
Cold & $70$  & $1.7\times 10^6$ & $1\times 10^3$ & $2.7\times10^5$ \\
\hline
Hot & $9$ & $9\times 10^4$ & $4\times 10^2$ & $0$ \\
\hline \hline
\end{tabular}
\caption{\label{times} Time-scales under Poynting-Robertson drag, collisions
and sublimation. At 70\,AU~: $\langle a^2\rangle\simeq 5\,a\dma{min}^2$,
$a\dma{eq} \simeq 0.26\times a$ (model \textbf{\#13}) and $T\dma{g}\simeq 90\,$K
(see Figure \ref{temp}).
At 9\,AU~: $\sigma_0\simeq0.5$\,grain.cm$^{-2}$,
$\langle a^2\rangle=a_0^2$ with $a_0\simeq450\,\mu$m,
$\rho\dma{g}=0.071$\,g.cm$^{-3}$ and H$_2$O ice is
sublimated ($a\dma{eq}=0$, see Figure \ref{dSub120K}). For both
amorphous grains with $a\dma{min}\!=\!10$\,$\mu$m and crystalline grains with
$a_0\!\simeq\!450$\,$\mu$m, we find $\langle Q\dma{abs}\rangle$ very close to $1$.}
\end{center}
\end{table}

Collisions are then the dominant process in the cold annulus.
The smallest grains produced by collisions in very short time-scales are not
efficiently removed from the ring by Poyn\-ting-Robertson drag. Only radiation
pressure is then able to explain the minimum grain size ($10\,\mu$m) derived
from the SED fitting.
Also, this is consistent with the fact that this minimum grain size is very
close to the blow-out size limit. Of course, this argument applies only if the
gas does not play an important role, as suggested in a previous subsection
(\ref{gdratio}).
At 70\,AU from the star, the H$_2$O ice within the smallest grains
is sublimated in a short time-scale in comparison
with the star age. This is consistent with the little percentage of H$_2$O ice
required to fit the 20 -- 25\,$\mu$m SED (subsection \ref{cold}).

The hot dust population is poorly constrained; the time-scales in Table
\ref{times} have then to be taken with care. The life time of the $450\,\mu$m
crystalline grains under Poynting-Robertson drag is long compared to the
collision time-scale, and even $120\,\mu$m crystalline grains with
$\beta\dma{pr}\simeq 0.5$ have $t\dma{PR}$ about 4 times larger than
$t\dma{coll}$. Also, collisions and radiation pressure are probably the
two main processes which shape the hot dust distribution.
\subsection{Ring Dynamics  and confinement}
\subsubsection{Collision dynamics and Mass loss}
\label{Mloss}
We here consider the implications of the model parameters for the ring
centred at $70$\,AU given in Table \ref{3.5OUTERSED}. As representative
example we consider model \textbf{\#13}. This has a total grain surface density
$\sigma_{0} = 1.79\times 10^3\,$grains.cm$^{-2}$. For the assumed size
distribution ($\propto a^{-3.5}$), the mean collision time for the
smallest grains with $a=a\dma{min}\simeq 10\,\mu$m is approximately
$t\dma{coll}(a\dma{min})\simeq 2.1\times 10^3\,$yr.
As these grains are at the radiation pressure blow out limit and the
majority of grains have the minimum size, the minimum mass
loss rate is given by~:
\begin{eqnarray*}
{\dot M} = \frac{M\dma{d}(a\dma{min})}{t\dma{coll}(a\dma{min})}=
{4\pi a\dma{min}^3\rho\dma{g}n(a\dma{min})\over 3\,t\dma{coll}(a\dma{min})}
\sigma_0 A
\end{eqnarray*}
where $A=1.7\times 10^{30}\,$cm$^2$ is the area of the annulus.
For model \textbf{\#13} this corresponds to ${\dot M} \simeq 2.5\times 10^{-6}
M_{\oplus}$yr$^{-1}$. Thus the dust mass reservoir required to supply
this mass loss over the age of the system is $\sim 20 M_{\oplus},$
which corresponds to a surface density in solids $\Sigma\dma{gr} \simeq
0.07$\,g.cm$^{-2}$.

It is of interest to note that if we assume a gas to dust ratio of
$100$ in the primordial nebulae, this corresponds to a surface density
$\Sigma \simeq 7$ for the original protostellar disk. This surface density is
consistent with that expected at $70\,$AU for standard disk models
corresponding to a minimum mass solar nebula (e.g. Papaloizou \&\
Terquem 1999).

If the surface density $\Sigma\dma{gr}$ resides predominantly in objects
with $a=a\dma{max} \sim 1$\,m, their collision time would be $t\dma{coll} \sim
2a\rho\dma{g}/(3\Omega \Sigma\dma{gr}) \simeq 8.6\times 10^4\,$yr.
This is still shorter than the age of the system. With a ring aspect ratio of
$0.1$ for instance, collisions are expected to occur with a relative velocity
$\sim 0.5\,$km.s$^{-1}$, and are likely to be destructive. In order to have a
collisional lifetime greater than the star age, the objects would have to
have $a$ larger than about $100$\,m.
A population of such larger objects providing a mass
source for those with $a \sim 1\,$m, may exist but they would not be
detectable. In that case the disk mass could be significantly larger
than $4\,M_{\oplus}$.

\subsubsection{Formation, vertical height and possible presence of larger
bodies}
In the standard model of planetary formation \citep{wei93},
objects in the size range $1\,$m -- $1\,$km are formed by accumulation of
smaller objects through collisions when there was still gas present.
The presence of gas drag, effective enough here with $\Sigma \sim
7\,$g.cm$^{-2}$, would be essential to damp eccentricity and inclination so
collisions occur with small relative velocity.  However, the
collision timescales would be the same as estimated above.  Thus it
is reasonable that objects as large as $100\,$m could be formed if the
gas residence time was similar to the present age of the system.

When significant amount of gas was present during earlier
phases, the disk would be expected to be thin because of gas drag
\citep{wei93} but this could have been thickenned due to gravitational
perturbations from large external objects.
Unfortunately the observations are not yet able to provide contraints on the
vertical thickness, but this does not affect the discussion
about mass loss rates or collision times because these are independent of
the vertical thickness.
\subsubsection{Radial confinement}
The fact that the mass distribution in the ring is tightly radially
confined to a width of about $0.2$ times the mean radius,
which may be comparable to its vertical thickness,
gravitational confinement mechanisms may operate to sculpt its form.
As it is likely that many
collisions are required to reduce a metre sized object to fragments
small enough to be blown out by radiation pressure, if the 
the vertical thickness is as much as $0.2r$, the ring should be
significantly wider without a confinement mechanism.
Even if the ring is very thin
as it might have been immediately post formation
so that little collisional spreading occurs, a mechanism
is still required to explain why the protostellar gaseous
disk was truncated at around 70\,AU.

Confinement may occur through gravitational interaction with external
perturbers. This situation is familiar in the case of planetary rings
\citep{gol82} where the interaction with external
orbiting satellites provides confinement.  It also occurs for
accretion disks in close binary systems where interaction with a
binary companion truncates the disk \citep{pap77}.  The
presence of a binary companion in the system discussed here is
suggestive that it could play a role in determining the location of
the outer boundary to the disk as well as pumping up inclinations
after gas dispersal. Note also that an exterior
as yet undetected giant planet might provide a similar role.

The binary companion \hrb\ is seen at a projected distance of $515\,$AU from
the central star whereas the outer disk radius is $70\,$AU. Thus if it
is to play any role in confinement the orbit should be eccentric. As
significant interaction with the ring requires the pericentre distance
to be less than three times the ring radius, the minimum value for the
eccentricity is thus $e = 0.4$, although it could be significantly
larger on account of projection effects.

Disk edge truncation can be effective in the presence of binary
companions with significant mass ratio in circular or eccentric
orbits. In the case of very eccentric orbits, as may well be the case
here, the truncation process is most easily modelled as a succession
of encounters which occur at each pericentre passage
\citep{ost94,kor95,hall97}.  From these studies it is
clear that in order to be truncated, the disk radius should initially
exceed about one third of pericentre distance.  Material beyond this
radius is strongly perturbed and removed from the disk which develops
a truncated edge there. The latter phenomenon, which occurs because of
angular momentum transported from the disk to the companion, was seen
in the gas disk simulations of \citet{kor95} and the
free particle simulations of \citet{hall97} and is also expected for the
disks of the type considered here.
%
%
%
\section{Conclusion}
We propose a two dust populations model able to reproduce all the available
observations of the \hra\ \cs\ disk.
The first dust population, peaked at 70\,AU from the star, shapes like a narrow
annulus. Its small radial extension suggests a gravitational
confinement mechanism both outside and inside the peak at 70\,AU.
This ring is made of amorphous grains quite similar to those found in the
interstellar medium.
Under the assumption of a collisionnal grain size distribution, grains
larger than $10\,\mu$m to meter sized bodies are required to fit the data.
Collisions, occuring in short time scales, are expected to produce smallest
grains down to the blow out size limit ($\sim 10\,\mu$m).
This conclusion applies because of the gas to dust ratio smaller than 1
inferred. The second dust population at 9-10\,AU from the star is made
of comet-like grains. Higher signal to noise ratio measurements
and resolved images at $\lambda\sim 10\,\mu$m would allow to better constrain
this dust population (spatial extension, grain size range) and its dynamics.

The dust surrounding \hra\ probably shows an intermediate stage for the
\cs\ matter between pre-main sequence stars and evolved ones like \bp,
Fomalhaut, Vega, 55\,Cnc \ldots
The presence of a companion (\hrb) is also an interesting peculiarity among
the Vega-like stars.
As a large number of field stars are members of binary (or more) systems,
\hra\ does not represent a marginal case.
Therefore, the dynamics of the disk has to be further studied, in terms of
chemical evolution of the grains and dust distribution to better understand
the steps which lead or not to planetary systems formation.
\begin{acknowledgements}
We wish to thank G. Schneider and J.S. Greaves for providing data prior to
publication and S. Warren for providing water ice optical constants.
We also thank Herv\'e Beust for helpfull discussions
on the dynamics of the system.
All the computations presented in this paper were performed at the Service
Commun de Calcul Intensif de l'Observatoire de Grenoble (SCCI). 
\end{acknowledgements}

\end{document}